\documentclass{article}
\usepackage{setspace}
\usepackage{amsfonts}
\usepackage{amsmath}
\usepackage{graphicx}
\usepackage{amssymb}
\usepackage{amsmath}
\usepackage[pagewise]{lineno}
\usepackage{hyperref}

\input{tcilatex}
\begin{document}

\title{Turbulence phenomena for viscous fluids .
Vortices and instability}
\author{Mauro Fabrizio$^{1}$ \and $^{1}$ Department of Mathematics,
University of Bologna, Italy}
\maketitle

\begin{abstract}
Through the Ginzburg-Landau and the Navier-Stokes equations, we study
turbulence phenomena for viscous incompressible and compressible fluids by a
second order phase transition. For this model, the velocity is defined by
the sum of classical and whirling components. Moreover, the
laminar-turbulent transition is controlled by rotational effects of the
fluid. Hence, the thermodynamic compatibility of the differential system is
proved.

The same model is used to understand the origins of tornadoes and their
behavior and the birth of the vortices resulting from the fall of water in a
vertical tube. Finally, we demonstrate how the weak Coriolis force is able
to change the rotation direction of the vortices by modifying the minima of
the Ginzburg-Landau equation. Hence, we conclude the paper with the
differential system describing the water vorticity and its thermodynamic
compatibility.
\end{abstract}


\section{Introduction}

Turbulence phenomena in a viscous fluid is from a long time a controversial
problem for the study and modeling of the transition from laminar flow to
turbulent behavior\cite{Ar-St},\cite{Ch},\cite{FABE},\cite{Ru},\cite{St},%
\cite{Wi}. Many papers have tried modeling using only the classical Navier
Stokes equations. In other works, the turbulence is studied by a stochastic
cascade model. More recently, turbulence is described by a phase transition,
but not always with a suitable connection with the Navier Stokes equations.
In a previous work \cite{F2}, we have presented a transition model involving
a system, in which the Navier Stokes equations are coupled with the
Ginzburg-Landau equation \cite{F1},\cite{landau},\cite{Lifs. Pit} and where
the transition is described by a phase field $\varphi $\ and controlled not
by the velocity $v$ of the fluid, but by the function $\left \vert \nabla
\times \mathbf{v}\right \vert $. This choice is motivated by the results
presented in many articles \cite{Go}, \cite{Hu},\cite{St}, in which it is
observed that the roughness of the walls or the obstacles inside a channel
can anticipate the transition to turbulence, because they produce vortices.
Indeed, it seems to be not consistent to believe that a laminar flow in a
pipe, consisting only of perfectly parallel velocities, can be transformed
into a turbulent flow, when the velocity exceeds a given value predicted by
the Reynolds number. On the other hand, it seems reasonable to assume that
an appropriate (even small) disorder flow is needed for the transition.
Moreover, the classical view point by the use of Reynolds number \emph{R}
does not have local features, because depends on the duct radius and on mean
velocity so that cannot be considered in Navier-Stokes differential
equations. Finally, the transition is not well defined by \emph{R}, because
it work in a large interval of values of \emph{R}.

In phase transitions, the material can assume different structures.
Furthermore, the laminar and turbulence phases are characterized by two
different internal structures, as liquid and vapor. In recent years, we can
find in the literature numerous papers in which turbulence is studied as a
phase transition (see also\textbf{\ }$\cite{Ru},\cite{St},\cite{Zu})$

In the paper we also want to emphasize, that the model proposed will be
appropriate to describe the instability evident in the phenomena of
turbulence transitions. It would appear that this phenomenon has not been
well considered in the literature. In this paper, we recall the model of a
viscous incompressible fluid studied in \cite{F2} with some corrections and
improvements. In particular, we introduce a small (but important)
modification, because now the phase is in the set $\varphi \in (-1,1)$,
unlike previous work, where the phase was in the set $\varphi \in \left[
0,1\right) $. By this change, the instability effect is now described
through the bifurcation, which is manifest in the Landau potential \cite%
{landau} \cite{lanlif}, when the transition triggers. Because in such a case
the potential goes from a minimum to a double well. So otherwise of $\cite%
{F2}$, this effect and its instability generetes the turbulence.

Hence, in the paper we present an extension to compressible viscous fluids
and its thermodynamic compatibility. For these fluids we suppose the
threshold, that identifies the transition, is given by a suitable value of $%
\left \vert \nabla \times \mathbf{v}\right \vert $.

Moreover, there is a great similarity between turbulence phenomena in
viscous fluids and superfluids in Helium II (see \cite{Ar-St},\cite{F11},%
\cite{Kapit},\cite{landau},\cite{Lifs. Pit},\cite{Lond1},\cite{Mend1},\cite%
{Tisz}) and superconductivity \cite{G-L}.

Finally, as a consequence of this model, it follows that the differential
system can be well-posed problem only if we are in the laminar phase.
Otherwise, when we are in a turbulent flow regime, the instability of the
model makes the system not well posed.

In the last part of the paper, we consider phenomena for which the
transition does not produce turbulent effects, because the vortices have
ample dimensions as in tornadoes, or in the behavior of water disappearing
down the hole of a sink. To describe these new effects, we consider the same
differential systems, but with a different symmetry of Ginzburg-Landau
potentials. \ Therefore for these phenomena, the weak Coriolis force has an
important role, that is not to naturally generate the vortices, but to
influence indirectly the direction of rotation, because it is able to modify
the minima of the Ginzburg-Landau potential. Finally for this problems, we
study the differential system of a viscous incompressible fluid described by
a suitable connection between Navier-Stokes and Ginzburg-Landau equations.
As a consequence of this system, \ we obtain a particular theorem for an
extension of Navier-Stokes equations.

\section{Incompressible fluids, turbulence and unstable behavior}

In this paper, the mathematical model proposed for a turbulent flow in a
viscous fluid is confined in a smooth domain $\Omega \subset IR^{3}~$Hence,
we suppose the phenomenon as a consequence of a phase transition, such that
the transfer from laminar to turbulent flow is controlled by a phase field
(or order parameter) $\varphi \in \left( -1,1\right) $.

Following $\left[ 1\right] $, the velocity\ $\mathbf{v}$ of the fluid is
composed of the normal velocity $\mathbf{v}_{n}$ and the rotational
component $\mathbf{v}_{s}$. Moreover, for $\varphi \neq 0$ the
incompressible fluid velocities $\mathbf{v}_{n}$ and $\mathbf{v}_{s}$ are
related by the constraints%
\begin{equation}
\varphi (x,t)\mathbf{v}_{s}(x,t)\mathbf{=}\nu \mathbf{\nabla }\times \varphi
(x,t)\mathbf{v}_{n}(x,t)~,~~\nabla \cdot \mathbf{v}_{n}(x,t)=0  \label{1}
\end{equation}%
with $\nu $ a scalar and positive coefficient. Hence, we assume the velocity 
$\mathbf{v}$ to be given by the following relationship between $v_{n}$ and $%
v_{s}$ 
\begin{equation}
\mathbf{v}(x,t)=\mathbf{v}_{n}(x,t)+\varphi (x,t)\mathbf{v}_{s}(x,t)
\label{2}
\end{equation}%
Moreover, the phase $\varphi (x,t)~$satisfies the Ginzburg-Landau equation 
\begin{equation}
\rho _{0}\frac{d}{dt}\varphi (x,t)=\nabla \cdot L(x)\nabla \varphi
(x,t)-NF^{\prime }(\varphi (x,t))+\frac{\alpha }{2}G^{\prime }(\varphi (x,t))%
\mathbf{v}_{s}^{2}(x,t)  \label{3}
\end{equation}%
where $\rho _{0}>0$ is the density of the incompressible fluid, $\alpha >0$
, $L>0$ which are suitable parameters to the system structure\ and the dot $%
\cdot $~denotes the material time derivative, while $\nabla \cdot $ the
divergence operator. Moreover, the coefficient $N(\mu )$ can be a function
of the viscosity $\mu $. In addition, the two potentials $F$ and $G,$ are
such that $F$ is given by a parabolic function with minimum for $\varphi =0$
and $G$ by a function with a two well potential and maximum in $\varphi =0.$
As \ an example, we propose for $F$ and $G$ the functions 
\begin{equation}
F(\varphi )=\frac{\varphi ^{2}}{2},~~~~~G(\varphi )=\frac{\varphi ^{2}}{2}-%
\frac{\varphi ^{4}}{4},~~\text{with}~\  \varphi \in \left( -1,1\right)
\label{4}
\end{equation}

Finally, the velocity $\mathbf{v}_{n}$ satisfies a modified Navier-Stokes
equations%
\begin{equation}
\rho _{0}\frac{d}{dt}\mathbf{v}_{n}=-\nabla p-\mu \nabla \times \nabla
\times \mathbf{v}_{n}-\nu \alpha \varphi \nabla \times \varphi
^{-1}G(\varphi )\mathbf{\dot{v}}_{s}+\rho _{0}\mathbf{f}  \label{6}
\end{equation}%
where $p$ is the pressure, and $\mathbf{f}$ the body forces, while as in (%
\ref{3}) the material derivative $\frac{d}{dt}\mathbf{v}_{n}=\frac{\partial 
}{\partial t}\mathbf{v}_{n}+(\mathbf{v}_{n}\cdot \nabla \mathbf{v}_{n})$.
Finally, the new term $\nu \alpha \varphi \nabla \times \varphi
^{-1}G(\varphi )\mathbf{\dot{v}}_{s}$

Now, we introduce the boundary conditions on the smooth domain $\Omega \in
IR^{3}$ related with the system (\ref{3}-\ref{6}) on two domains $\partial
\Omega _{1}\neq 0$ and $\partial \Omega _{2}$

\begin{eqnarray}
\left. \nabla \times \mathbf{v}_{n}\times \mathbf{n}\right \vert _{\partial
\Omega _{1}} &=&0~,~\left. \mathbf{v}_{n}\times \mathbf{n}\right \vert
_{\partial \Omega _{2}}=0  \label{6aa} \\
&&  \notag \\
\left. \nabla \varphi \cdot \mathbf{n}\right \vert _{\partial \Omega } &=&0
\label{6b}
\end{eqnarray}%
these boundary conditions are supposed homogeneous only as a possible choice.

Per lo studio della evoluzione del sistema \`{e} cruciale il comportamento
del potenziale $W(\varphi ,\nabla \varphi )$ defined by%
\begin{equation*}
W(\varphi ,\nabla \varphi )=F(\varphi (x,t))-\frac{\alpha }{2N}\mathbf{v}%
_{s}^{2}(x,t)G(\varphi (x,t))
\end{equation*}

So, we have that the laminar-turbulent transition occurs when $(\frac{\alpha 
}{2N}v_{s}^{2}-1)$ changes sign. Indeed, for $\frac{\alpha }{2N}v_{s}^{2}>1$%
, we are in turbulent phase. Otherwise, if $\frac{\alpha }{2N}v_{s}^{2}<1$
we are in the situation of a laminar flow.

\begin{center}
\includegraphics[scale=0.4]{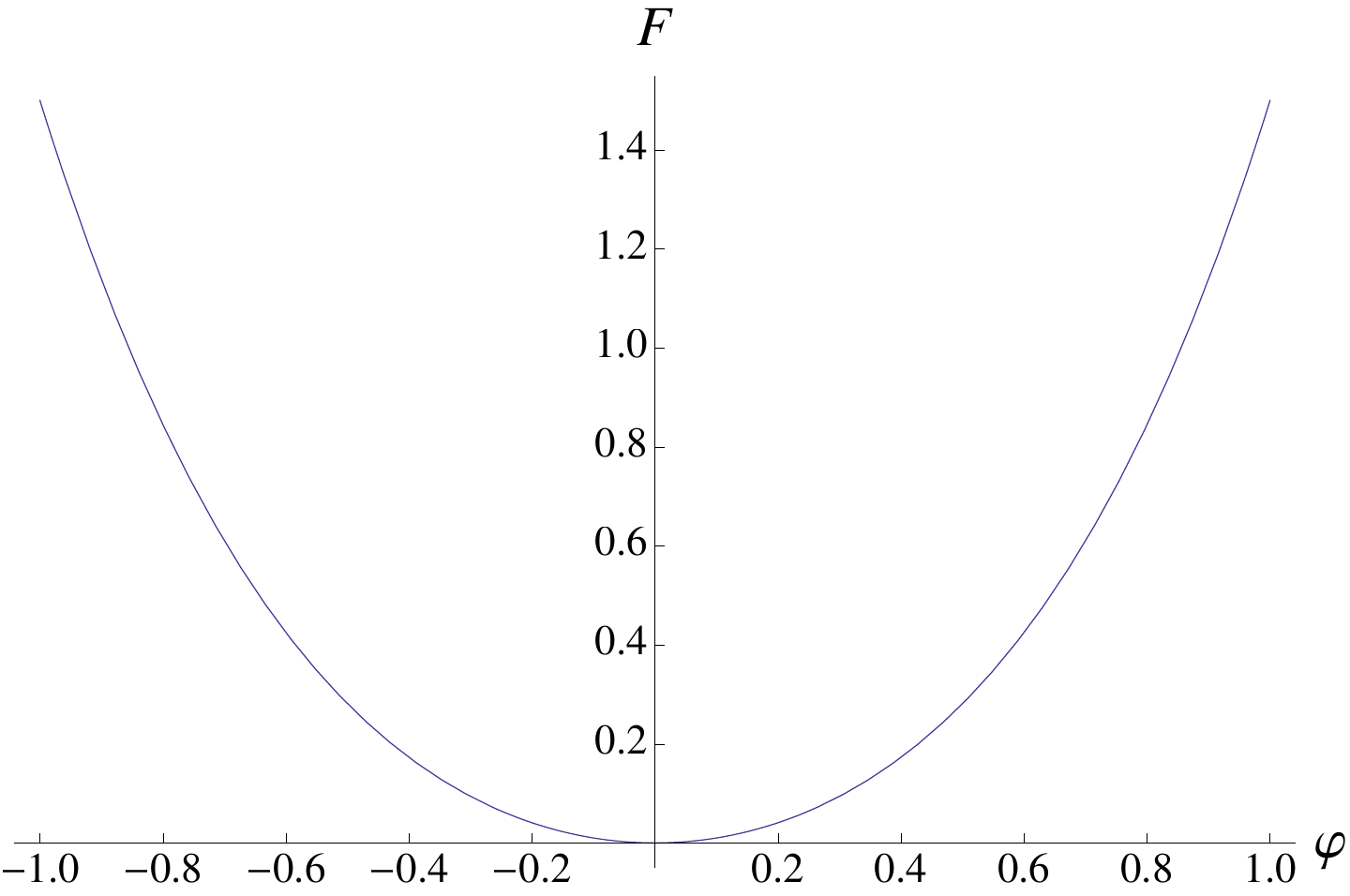} %
\includegraphics[scale=0.4]{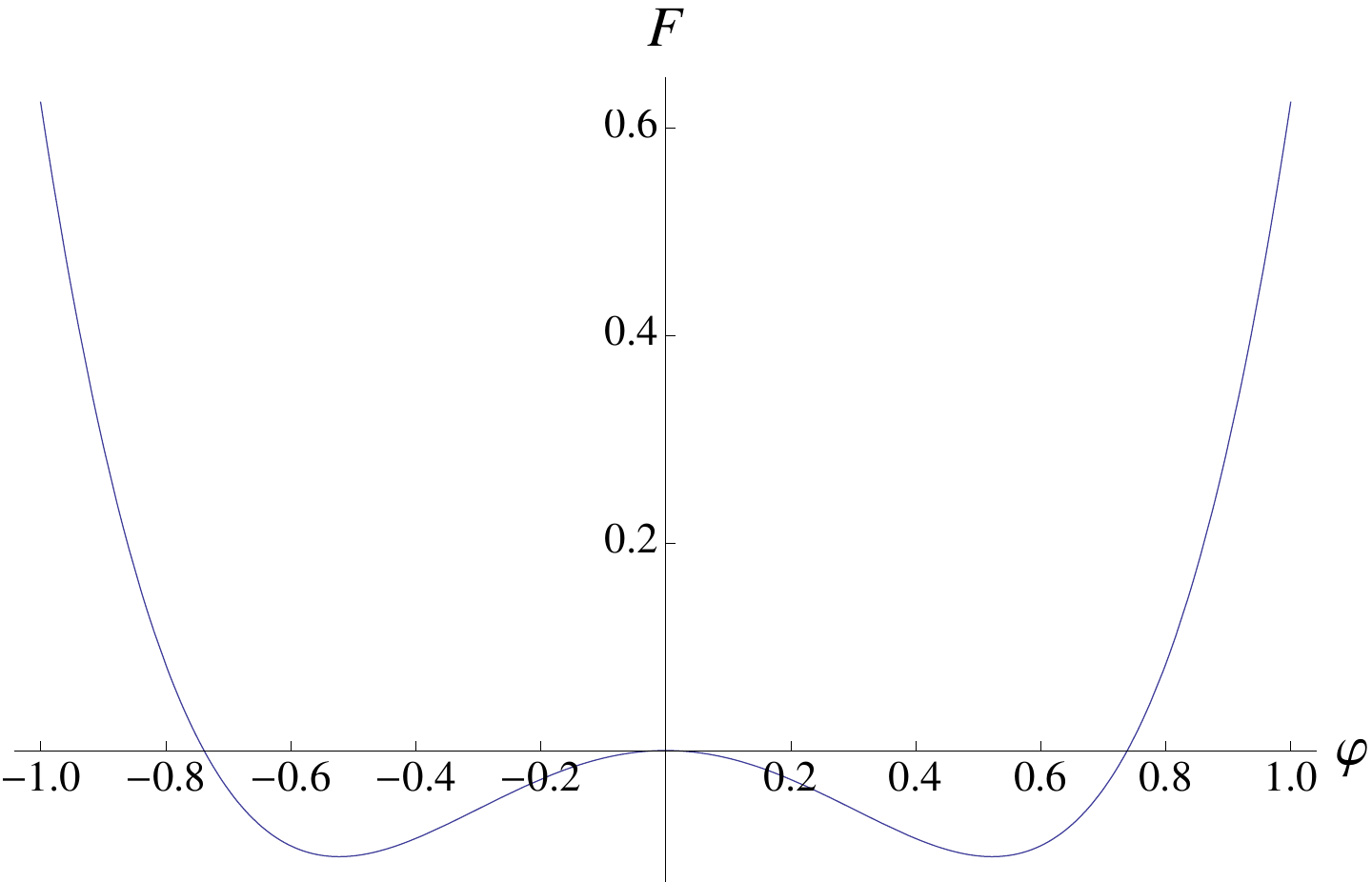}

Fig.1 - The two pictures describe the potential $W(\varphi )$ of (\ref{4})
the first and the potential $W(\varphi )$ the second.
\end{center}

The turbulence behavior is related to the double well function of$~W(\varphi
)$ $~$of (\ref{4}), because the two minima are at the same height. So that,
in such a case the system is unable to select between the two minima. So in
any point, we can have a different value of the minimum. So this instability
produces a turbulence behavior of the fluid.

Even if our system can be unstable, we \ are, in any case, always able to
obtain the compatibility of the system with Thermodynamics, which for
isothermal processes assumes the form of a Dissipation Principle \cite{F1},%
\cite{frem},\cite{Fr-Gur}:

There exists a kinetic energy $T=$ $\rho _{0}\mathbf{v}_{n}^{2}\ $and a
state function $\psi (\varphi ,\nabla \varphi ),$~called internal energy,
such that:%
\begin{equation}
\frac{d}{dt}T+\rho _{0}\frac{d}{dt}\Psi (\varphi ,\nabla \varphi )\leq \frac{%
1}{2}\frac{d}{dt}(\rho _{0}\mathbf{v}_{n}^{2}+\alpha G(\varphi )\mathbf{v}%
_{s}^{2})+  \label{8}
\end{equation}%
\begin{equation*}
+\mu (\nabla \times \mathbf{v)}^{2}+\rho _{0}\dot{\varphi}^{2}+\frac{1}{2}%
\frac{d}{dt}(L(\nabla \varphi )^{2}+2N\ F(\varphi ))
\end{equation*}

From the system (\ref{3}-\ref{6}), with the restriction (\ref{1}-\ref{2})
and the boundary conditions (\ref{6aa}-\ref{6b}), we obtain the internal
mechanical and structural power, respectively%
\begin{equation*}
P_{m}^{i}=\mu (\nabla \times \mathbf{v}_{n})^{2}
\end{equation*}%
\begin{equation*}
P_{s}^{i}=\rho _{0}\dot{\varphi}^{2}+(\frac{L}{2}(\nabla \varphi
)^{2})^{\cdot }+N~\dot{F}(\varphi )
\end{equation*}%
while the kinetic energy $T$ is given by%
\begin{equation}
T=\frac{1}{2}(\rho _{0}\mathbf{v}_{n}^{2}+\alpha G(\varphi )\mathbf{v}%
_{s}^{2})  \label{9}
\end{equation}

Finally, by Dissipation Principle%
\begin{equation*}
\rho _{0}\dot{\psi}(\varphi ,\nabla \varphi )\leq P_{m}^{i}+P_{s}^{i}
\end{equation*}%
we obtain the inequality (\ref{8}) with free energy 
\begin{equation}
\psi (\varphi ,\nabla \varphi )=\frac{1}{2\rho _{0}}(L(\nabla \varphi
)^{2}+2N\ F(\varphi )).  \label{9a}
\end{equation}%
while the dissipation $D$ is given by%
\begin{equation*}
D=\mu (\nabla \times \mathbf{v}_{n})^{2}+\rho _{0}\dot{\varphi}^{2}
\end{equation*}

\section{Maximum theorem}

Now, we study a maximum theorem for differential problem, of the previous
section, that we rewrite by the system%
\begin{eqnarray}
\gamma \frac{d}{dt}\varphi &=&\nabla \cdot L(x)\nabla \varphi -NF^{\prime })+%
\frac{\alpha }{2}G^{\prime }(\varphi )\mathbf{v}_{s}^{2}  \label{22aa} \\
&&  \notag \\
\rho _{0}\frac{d}{dt}\mathbf{v}_{n} &=&-\nabla p-\mu \nabla \times \nabla
\times \mathbf{v}_{n}-  \notag \\
&&-\nu \alpha \varphi \nabla \times \varphi ^{-1}G(\varphi )\mathbf{\dot{v}}%
_{s}+\rho _{0}\mathbf{f}  \label{22b} \\
&&  \notag \\
\varphi (x,t)\mathbf{v}_{s}(x,t) &\mathbf{=}&\nu \mathbf{\nabla }\times
\varphi (x,t)\mathbf{v}_{n}(x,t)\text{ },~\  \nabla \cdot \mathbf{v}_{n}=0
\label{22c}
\end{eqnarray}%
with the boundary conditions (\ref{6aa}) and (\ref{6b}) and the initial
conditions%
\begin{equation}
\varphi (x,0)=\varphi _{0}(x)~,~\ ~~\mathbf{v}_{n}(x,0)=\mathbf{v}%
_{n}^{0}(x)~,~\ x\in \Omega  \label{24b}
\end{equation}

Because with $\varphi =0$ we denote the laminar phase, while $-1<\varphi <1$
represents the non-laminar phase. So we have to prove that we cannot to have 
$\varphi ^{2}\neq \left[ 0,1\right) .$In other words, we prove the following
maximum theorem:

\textbf{Theorem}. Any solution of the problem (\ref{22aa})-(\ref{22c}), with
initial conditions (\ref{24b}), where $\varphi _{0}(x)\in \left( -1,1\right)
,$ is such that $\varphi (x,t)\in \left( -1,1\right) $ \ almost everywhere
in $\Omega \times (0,\infty ).$

Proof. We multiply the equation (\ref{22aa}) by $\varphi $. Then without
loss of generality, we suppose for this prooof $L(x)=\rho _{0}=1.$ Then, by (%
\ref{24b}) we obtain 
\begin{eqnarray}
\gamma \frac{d}{dt}\frac{\varphi ^{2}}{2} &\leq &\varphi \nabla ^{2}\varphi
+\alpha \mathbf{v}_{s}^{2}(\varphi ^{4}-\varphi ^{2})=  \label{25} \\
&&  \notag \\
&&-\frac{1}{2}\nabla ^{2}\varphi ^{2}-(\nabla \varphi )^{2}-\alpha ^{2}%
\mathbf{v}_{s}^{2}\varphi ^{2}(\varphi ^{2}-1)  \notag
\end{eqnarray}%
from which we have 
\begin{equation}
\frac{\gamma }{2}\frac{d}{dt}(\varphi ^{2}-1)\leq -\frac{1}{2}\nabla
^{2}(\varphi ^{2}-1)+\alpha ^{2}\mathbf{v}_{s}^{2}\varphi ^{2}(\varphi
^{2}-1)  \label{26}
\end{equation}%
hence, we introduce the function%
\begin{equation}
(\varphi ^{2}-1)_{+}=\left \{ 
\begin{array}{c}
0~~,\  \  \  \varphi \in (-1,1)~\  \  \  \  \  \  \\ 
\varphi ^{2}-1~\ ,~~\varphi \notin \left( -1,1\right)%
\end{array}%
\right.  \label{27}
\end{equation}%
multiplying (\ref{26}) by $(\varphi ^{2}-1)_{+}$ and\ after an integration
on $\Omega ,$ we have the inequality%
\begin{equation}
\begin{split}
\frac{\gamma }{4}\frac{d}{dt}\int_{\Omega }(\varphi ^{2}-1)_{+}^{2}~dx& +%
\frac{1}{4}\int_{\Omega }(\nabla (\varphi ^{2}-1))_{+}^{2} \\
& +\alpha ^{2}\mathbf{v}_{s}^{2}\varphi ^{2}(\varphi ^{2}-1)_{+}^{2})dv\leq 0
\end{split}
\label{28}
\end{equation}%
hence, by the divergence theorem and boundary condition (\ref{23})$_{1}$, we
have 
\begin{equation}
\frac{\gamma }{4}\frac{d}{dt}\int_{\Omega }(\varphi ^{2}-1)_{+}^{2}~dx+\frac{%
1}{4}\int_{\Omega }\alpha ^{2}\mathbf{v}_{s}^{2}\varphi ^{2}(\varphi
^{2}-1)_{+}^{2}dv\leq 0  \label{29}
\end{equation}%
Finally, after a time integration on $\left[ 0,T\right] $ and by the initial
condition (\ref{24})$_{1}$, we obtain%
\begin{equation}
\begin{split}
& \frac{\gamma }{4}\int_{\Omega }(\varphi ^{2}(x,t)-1)_{+}^{2}~dx+ \\
& \quad \frac{1}{4}\int_{0}^{T}\int_{\Omega }\alpha ^{2}\mathbf{v}%
_{s}^{2}\varphi ^{2}(x,t)(\varphi ^{2}(x,t)-1)_{+}^{2}dvdt\leq 0
\end{split}%
\end{equation}%
from which $(\varphi ^{2}(x,t)-1)_{+}=0$. for any $t\geq 0$. Then, because $%
\varphi _{0}(x)\in \left( -1,1\right) $ the theorem is proved.

\section{Generalized Navier-Stokes and Ginzburg-Landau equations for
turbulent compressible fluids}

As in the previous section, we suppose the velocity of a compressible fluid
of density $\rho $ composed of a normal velocity $v_{n}$ together with the
rotational component $v_{s}$ of the same equation (\ref{2}), while the
equation (\ref{1}) is replaced by the following 
\begin{equation}
\varphi (x,t)\mathbf{v}_{s}(x,t)=\tilde{\lambda}\nabla \times \varphi (x,t)%
\mathbf{v}_{n}(x,t)  \label{11}
\end{equation}%
where $\tilde{\lambda}$ is a new positive coefficient. Hence, from (\ref{11}%
) we \ obtain%
\begin{equation}
\nabla \cdot (\varphi \mathbf{v}_{s})=0  \label{11a}
\end{equation}%
then from the continuity equation and utilizing the equation (\ref{2}) we
obtain%
\begin{equation}
\frac{\partial \rho }{\partial t}=-\nabla \cdot (\rho \mathbf{v)=-}\nabla
\cdot (\rho \mathbf{v}_{n}\mathbf{)-}\rho \mathbf{v}_{s}\cdot \nabla \varphi
\label{12}
\end{equation}%
Moreover, by (\ref{11a}) \ we have

\begin{equation*}
\frac{d}{dt}\rho =-\rho \nabla \cdot \mathbf{v=}-\rho \nabla \cdot (\mathbf{v%
}_{n}+\varphi \mathbf{v}_{s})=-\rho \nabla \cdot \mathbf{v}_{n}
\end{equation*}%
Therefore, we have the motion equation 
\begin{eqnarray}
\rho \mathbf{\dot{v}}_{n} &=&-\nabla p(\rho )+(2\mu +\lambda )\nabla \cdot
\nabla \mathbf{v}_{n}-\mu \nabla \times \nabla \times \mathbf{v}_{n}+
\label{13} \\
&&  \notag \\
&&+\alpha \varphi \nabla \times \varphi ^{-1}(G(\varphi )\mathbf{\dot{v}}%
_{s})+\rho \mathbf{f}  \notag
\end{eqnarray}%
where $\mu $~and $\gamma $ are positive constants, such that $\lambda \gamma
=\alpha $ and $\mathbf{f}$ is the body force.

Finally, as in the previous section, we consider the Ginzburg-Landau equation%
\begin{equation}
\rho \dot{\varphi}=\nabla \cdot \rho L\nabla \varphi -NF^{\prime }(\varphi )+%
\frac{\alpha }{2}\alpha G^{\prime }(\varphi )\mathbf{v}_{s}^{2}  \label{14}
\end{equation}%
where the coefficients $N,L$and the potentials $F(\varphi )$ and $G(\varphi
) $ are the same as in the previous section, defined by (\ref{4}).

Thus, the internal power $P_{m}^{i},P_{s}^{i}~$is given by%
\begin{equation}
P_{m}^{i}=p(\rho )\frac{\dot{\rho}}{\rho }+\mu (\nabla \times \mathbf{v}%
_{n})^{2}+(2\mu +\lambda )(\nabla \mathbf{v}_{n})^{2}  \label{14a}
\end{equation}

\QTP{Body Math}
\begin{equation}
P_{s}^{i}=\rho \dot{\varphi}^{2}+(\rho \frac{L}{2}(\nabla \varphi
)^{2})^{\cdot }+\rho N~\dot{F}(\varphi )  \label{14b}
\end{equation}

\bigskip While, the free energy $\psi (\varphi ,\nabla \varphi )\ $is
defined by 
\begin{equation}
\psi (\varphi ,\nabla \varphi )=\frac{L}{2}(\nabla \varphi )^{2}+NF(\varphi )
\label{14'}
\end{equation}%
and the kinetic energy $T$%
\begin{equation}
T(\rho ,\mathbf{v}_{n},\mathbf{v}_{s},\varphi )=\frac{1}{2}(\rho \mathbf{v}%
_{n}^{2}+\alpha G(\rho )\mathbf{v}_{s}^{2})  \label{14c}
\end{equation}

Hence, the free energy $\psi (\varphi ,\nabla \varphi )$~satifies the
inequality%
\begin{equation}
\rho \dot{\psi}(\varphi ,\nabla \varphi )\leq P_{m}^{i}+P_{s}^{i}=\mu
(\nabla \times \mathbf{v}_{n})^{2}+
\end{equation}%
\begin{equation*}
+(2\mu +\lambda )(\nabla \mathbf{v}_{n})^{2}+\rho \dot{\varphi}^{2}+\rho 
\frac{d}{dt}(\frac{L}{2}\nabla \varphi )^{2}+N~F(\varphi ))
\end{equation*}

\section{Tornadoes and ciclones.}

A model similar to one considered in the previous sections, can descrive the
vortices typical of tornadoes and cyclones, but also the water vorticity
observed when the water descends into a sink or more generally in a hole.
For those phenomena, as tornadoes, characterized by compressible viscous
fluids, we consider the equations (\ref{1}-\ref{6}) with a polynomial $%
\tilde{G}(\varphi )$ defined in 
\begin{equation}
\tilde{G}(\varphi )=\frac{\varphi ^{2}}{2}-\frac{\varphi ^{4}}{4}+b\left(
\varphi -\frac{\varphi ^{3}}{3}\right) ,~~\text{with}~\  \varphi ,b\in \left(
-1,1\right)  \label{16a}
\end{equation}%
such that the coefficient $b~$in (\ref{16a}) is related with the component
of Coriolis force, defined by 
\begin{equation}
b=\tau \mathbf{\omega \times v}_{r}\cdot \mathbf{t}  \label{17a}
\end{equation}%
where $\mathbf{t}$ is the tangent to the meridian, the vector $\mathbf{%
\omega }$ the angular velocity of the earth and $\mathbf{v}_{r}$ the
velocity relative to the terrestrial surface. Finally, the scalar $\tau $ is
a constant coefficient, related with the mass of the body.

So, the new potential $\tilde{W}$ is given by%
\begin{equation}
\tilde{W}(\varphi )=N(\mu )\tilde{F}(\varphi (x,t))-\frac{\alpha }{2}\tilde{G%
}(\varphi (x,t))(\mathbf{\nabla }\times \varphi (x,t)\mathbf{v}_{n}(x,t))^{2}
\label{17b}
\end{equation}%
then , according to the sign of $b$, we obtain the following two graphs for $%
\tilde{F}(\varphi )$

\begin{center}
\includegraphics[scale=0.4]{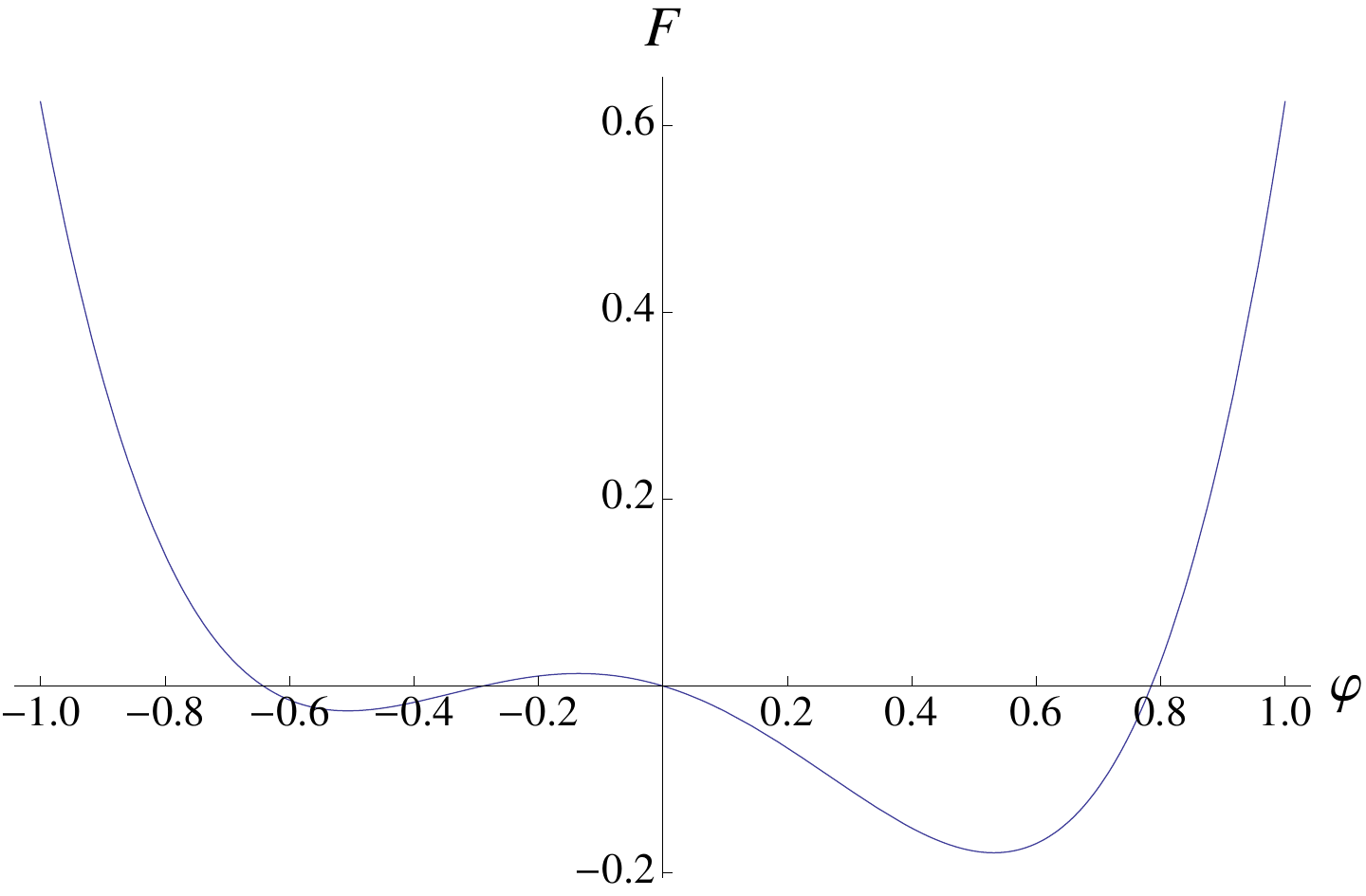} %
\includegraphics[scale=0.4]{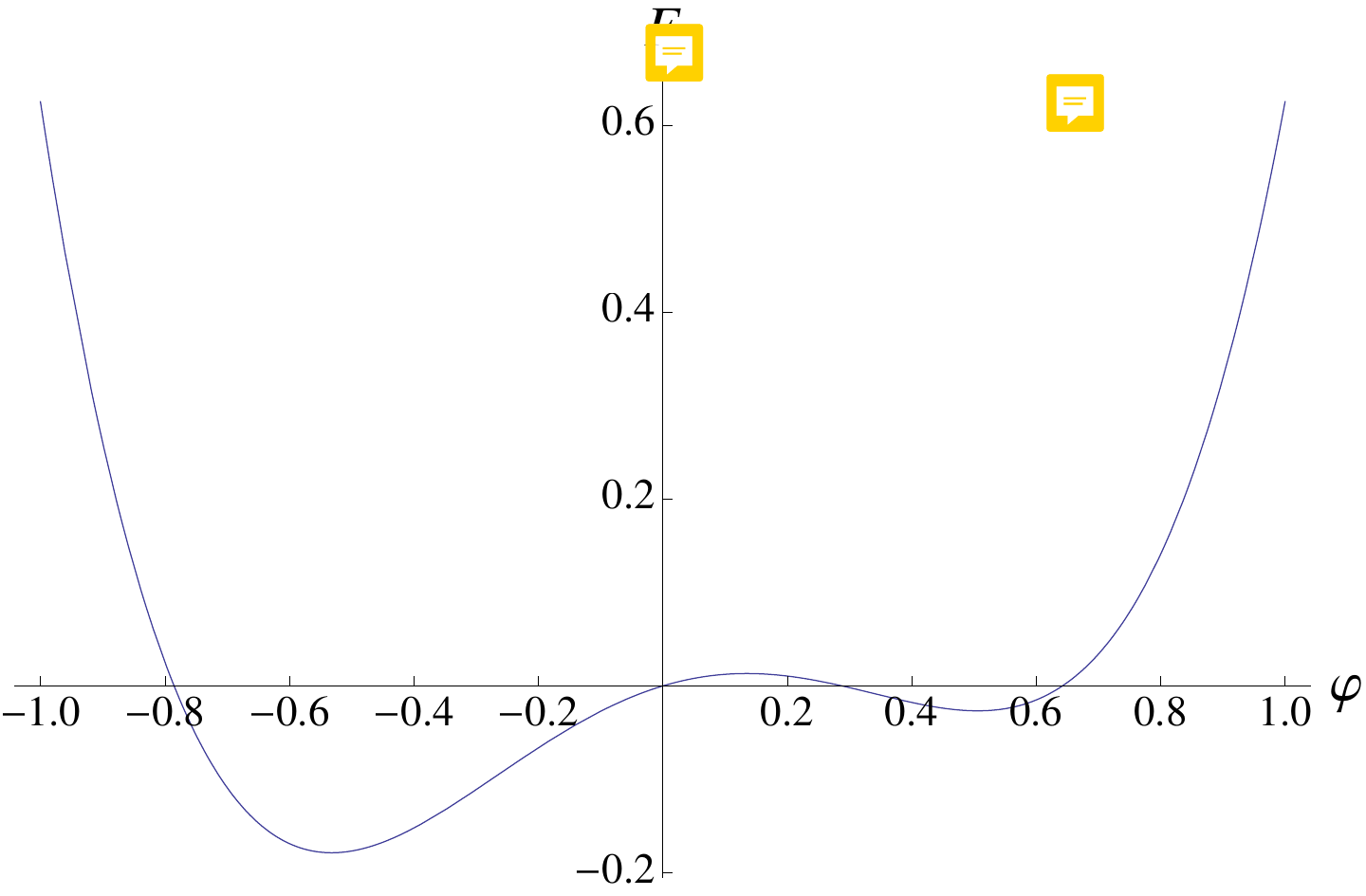}

Fig.2 - The two pictures are obtained with two different values of the
coefficients $b.~$In one we have $b>0$ and in the second $b<0.$
\end{center}

For tornados and cyclones we need to consider the system for a compressible
viscous fluid with the function $\tilde{G}(\varphi )~$defined in (\ref{16a}%
), So, the velocities $\mathbf{v}_{n}$ and $\mathbf{v}_{s}\mathbf{\ }$are
related by (\ref{11}), while $\mathbf{v}_{n},$ and $\varphi $ satisfy 
\begin{eqnarray}
\rho \mathbf{\dot{v}}_{n}(x,t) &=&-\nabla p(\rho (x,t))+\lambda \nabla
\nabla \cdot \mathbf{v}_{n}(x,t)+2\mu \nabla \cdot \nabla \mathbf{v}%
_{n}(x,t)+  \label{18} \\
&&  \notag \\
&&+\tilde{\nu}\varphi (x,t)\nabla \times \varphi ^{-1}(x,t)(\tilde{G}%
(\varphi (x,t))(\varphi (x,t)\mathbf{v}_{s}(x,t)\mathbf{)}^{\cdot })+\rho 
\mathbf{f}  \notag
\end{eqnarray}%
\begin{eqnarray}
&&\rho (x,t)\dot{\varphi}(x,t)=  \label{19} \\
&&  \notag \\
&=&\nabla \cdot \rho (x,t)L\nabla \varphi (x,t)-\rho NF^{\prime }(\varphi
(x,t))+\frac{\alpha }{2}\tilde{G}^{\prime }(\varphi (x,t))(\varphi (x,t)%
\mathbf{v}_{s}(x,t)\mathbf{)}^{2}  \notag
\end{eqnarray}

Now, we observe that the rotation movement of clouds that occur both in the
tornadoes and cyclones, is due to ascending hot air motions. When the flow
arrived in the atmosphere in contact with cold clouds, undergo a deviation
of the downward motion that causes a relapse of the flow on the earth. The
difference between tornadoes and cyclones is due to the motion of fluid. For
the tornadoes, the fall occurs inside the upward motion, for the cyclones
the flow of hot air due to the impact with the cold in altitude, tends to
widen and move away from the ascending flow. In both cases, the ascending
rectilinear motion, when is diverted, generates a component $\varphi \mathbf{%
v}_{s}$ in the equation (\ref{11}), that added to $\mathbf{v}_{n}$ produces
a whirling motion.To understand these vorticose phenomena and the difference
with turbulence, we have to observe the expression (\ref{16a}) of $\tilde{G}$%
, which compared to $G$ of (\ref{4})$_{2}$ presents the addend $b\left(
\varphi -\frac{\varphi ^{3}}{3}\right) .$The function $G(\varphi )$ is
always symmetrical in $\varphi $, as the function $W(\varphi )$. On the
contrary, $\tilde{G}$ is asymmetric, just like $\tilde{W}(\varphi )$.
Moreover, in the laminar phase the function $\tilde{W}(\varphi )$ always has
only one minimum in $\varphi =0.\ $Instead in the turbulent phase due to $%
\tilde{G}$, the potential $\tilde{W}(\varphi )$ lose the symmetry and has
two minima at different heights, depending on $b$. Finally, this model is
able to explain the different rotation observed in the two terrestrial
hemispheres. In fact, the minima of the potential $\tilde{W}(\varphi )$ are
inverted according to the sign of $b,$ related with the Coriolis force.

\section{Water vorticity}

A model similar to one considered in the previous sections can descrive the
water vorticity, observed when it descends into a sink or more generally in
a hole. For this problem related with incompressible viscous fluids, we use
the equations (\ref{1}-\ref{6}) with a similar polynomial $\tilde{G}(\varphi
)$ defined in (\ref{16a}) in order to describe the direction of the
vorticity depending on the hemisphere, we suppose again the coefficient $b$
defined as in (\ref{17a}). So, the potential $\tilde{W}$ is given by (\ref%
{17b})$.$

Now, we study the system representing a incompressible viscous fluid, with
the function $\tilde{G}(\varphi )~$defined in (\ref{16a}), So, the
velocities $\mathbf{v}_{n}$ and $\mathbf{v}_{s}\mathbf{\ }$are related by (%
\ref{11}), while $\mathbf{v}_{n},$ and $\varphi $ satisfy 
\begin{eqnarray}
&&\rho _{0}\mathbf{\dot{v}}_{n}(x,t)=-\nabla p(x,t)-\mu \nabla \times \nabla
\times \mathbf{v}_{n}(x,t)-  \label{18aa} \\
&&  \notag \\
&&-\nu \varphi \nabla \times \varphi ^{-1}(x,t)\tilde{G}(\varphi (x,t))%
\mathbf{\dot{v}}_{s}(x,t)+\rho _{0}\mathbf{f}(x,t)  \notag
\end{eqnarray}%
\begin{equation}
\rho _{0}\dot{\varphi}(x,t)=\nabla \cdot \rho _{0}L\nabla \varphi
(x,t)-NF^{\prime }(\varphi (x,t))+\frac{\nu }{2}\rho \tilde{G}^{\prime
}(\varphi (x,t))(\mathbf{v}_{s}(x,t)\mathbf{)}^{2}  \label{19a}
\end{equation}%
moreover, $\nabla \cdot \mathbf{v}_{n}=0$, $F(\varphi )=\frac{\varphi ^{2}}{2%
}\ $and with the conditions (\ref{11}), (\ref{11a}) and the boundary
conditions (\ref{6aa}) and (\ref{6b}).

So, for this problem if we multiply the equation (\ref{18aa}) for $\mathbf{v}%
_{n}$ and (\ref{19a}) for $\dot{\varphi},$ we have%
\begin{eqnarray}
&&\int_{\Omega }\rho _{0}\mathbf{\dot{v}}_{n}(x,t)\cdot \mathbf{v}%
_{n}(x,t)dx=\int_{\Omega }(-\mu \left( \nabla \times \mathbf{v}%
_{n}(x,t)\right) ^{2}-  \label{20} \\
&&  \notag \\
&&-\nu \tilde{G}(\varphi (x,t))\mathbf{\dot{v}}_{s}(x,t)\cdot \mathbf{v}%
_{s}(x,t)+\rho _{0}\mathbf{f}(x,t)\cdot \mathbf{v}_{n}(x,t))dx  \notag
\end{eqnarray}%
and 
\begin{eqnarray}
&&\int_{\Omega }\rho _{0}\dot{\varphi}^{2}(x,t)dx  \label{21} \\
&&  \notag \\
&=&\int_{\Omega }-\frac{\rho _{0}}{2}L\left( \left( \nabla \varphi
(x,t)\right) ^{2}\right) ^{\cdot }-N\dot{F}(\varphi (x,t))-\frac{\nu }{2}(%
\tilde{G})^{\cdot }(\varphi (x,t))(\mathbf{v}_{s}(x,t)\mathbf{)}^{2}  \notag
\end{eqnarray}%
from which by the boundary conditions (\ref{6aa}) and (\ref{6b}), we obtain
the energy balance law%
\begin{eqnarray}
&&\frac{d}{dt}\int_{\Omega }(\frac{1}{2}\rho _{0}\mathbf{v}_{n}^{2}(x,t)+%
\frac{\rho _{0}}{2}L\left( \left( \nabla \varphi (x,t)\right) ^{2}\right) +%
\frac{\nu }{2}\tilde{G}(\varphi (x,t))(\mathbf{v}_{s}(x,t)\mathbf{)}^{2})dx=
\label{22} \\
&&  \notag \\
&&-\int_{\Omega }(\frac{d}{dt}NF(\varphi (x,t))+\rho _{0}\dot{\varphi}%
^{2}(x,t)+\mu \left( \nabla \times \mathbf{v}_{n}(x,t)\right) ^{2}+\rho _{0}%
\mathbf{f}(x,t)\cdot \mathbf{v}_{n}(x,t))dx  \notag
\end{eqnarray}

Now, we study a week representetion of the equations (\ref{18aa}) and (\ref%
{19}) by the use of smooth functions of compact support $\boldsymbol{v}_{n}$
and $\phi ,$ which belong to the set $H^{1}(0,T;H_{0}^{1}(\Omega ))$ by the
following system\qquad 
\begin{eqnarray}
&&\int_{\Omega }\rho _{0}\mathbf{v}_{n}(x,t)\cdot \boldsymbol{\dot{v}}%
_{n}(x,t)dx=\int_{\Omega }(-\mu \nabla \times \mathbf{v}_{n}(x,t)\cdot
\nabla \times \boldsymbol{v}_{n}(x,t)-  \label{23} \\
&&  \notag \\
- &&\nu \tilde{G}(\varphi (x,t))\mathbf{\dot{v}}_{s}(x,t)\cdot \boldsymbol{v}%
_{s}(x,t)+\rho _{0}\mathbf{f}(x,t)\cdot \boldsymbol{v}_{n}(x,t))dx  \notag
\end{eqnarray}

and \ 
\begin{eqnarray}
&&\int_{\Omega }\rho _{0}\varphi (x,t)\dot{\phi}(x,t)dx=\int_{\Omega }-\frac{%
\rho _{0}}{2}L\nabla \varphi (x,t)\cdot \nabla \dot{\phi}(x,t)-  \label{24}
\\
&&  \notag \\
&&-NF^{\prime }(\varphi (x,t))\dot{\phi}(x,t)-\frac{\nu }{2}\tilde{G}%
^{\prime }(\varphi (x,t))\dot{\phi}(x,t)(\mathbf{v}_{s}(x,t)\mathbf{)}^{2} 
\notag
\end{eqnarray}

We can finich with particular phenomena for which we can suppose the phase
field $\varphi $ a constant function in $(x,t).$ Then, we work only with the
equation (\ref{23}), where the function $\varphi $ is a constant.

Then, under the condition

\begin{equation}
\nabla \cdot \mathbf{v}_{n}(x,t)=0  \label{24a}
\end{equation}%
the local form of the equation (\ref{23}) is given by%
\begin{equation}
\rho _{0}\frac{d}{dt}\mathbf{v}_{n}(x,t)=-\nabla p-\mu \nabla \times \nabla
\times \mathbf{v}_{n}(x,t)-  \label{25aa}
\end{equation}%
\begin{equation*}
-\tilde{\nu}\nabla \times \frac{d}{dt}\nabla \times \mathbf{v}_{n}(x,t)+\rho
_{0}\mathbf{f}(x,t)
\end{equation*}%
whose energy equation is represented by%
\begin{equation}
\frac{\rho _{0}}{2}\frac{d}{dt}\mathbf{v}_{n}^{2}(x,t)+\tilde{\nu}\frac{d}{dt%
}\left( \nabla \times \mathbf{v}_{n}(x,t)\right) ^{2}=  \label{26aa}
\end{equation}%
\begin{equation*}
=-\mu \left( \nabla \times \mathbf{v}_{n}(x,t)\right) ^{2}+\rho _{0}\mathbf{f%
}(x,t)\cdot \mathbf{v}_{n}(x,t)
\end{equation*}

Hence, in the space $H^{1}(0,T;L^{2}(\Omega ))\cap
L^{2}(0,T;H_{0}^{1}(\Omega )),$ we can to study the existence and uniqueness
of the system (\ref{24a}) (\ref{25aa}) under suitable smooth initial and
boundary conditions%
\begin{equation*}
I.C.\  \  \  \  \mathbf{v}_{n}(x,0)=\mathbf{v}_{0}(x),\  \  \  \nabla \times 
\mathbf{v}_{n}(x,0)=\nabla \times \mathbf{v}_{0}(x)
\end{equation*}

\begin{equation*}
B.C.\  \  \  \  \  \  \left. \mathbf{v}_{n}(x,t)\right \vert _{\partial \Omega }=0
\end{equation*}

\section{Conclusions}

The model studies a system composed of a modified Navier-Stokes and
Ginzburg-Landau equations. This system wants to describe the
laminar-turbulent transition in a viscous fluid. Where the transition is
controlled by the $curl\  \mathbf{v}$ and is activated when this coefficient
exceeds a fixed threshold. In this case, from the Ginzburg-Landau equation,
the phase $\varphi $ assumes values {}{}different from sero. Therefore, if
we have a potential with the same quote of the two minima, then turbulence
occurs due to the instability of the system.

Finally, using the same model, we will study the vorticity of water in a
tube and the birth of tornadoes and cyclones. For these phenomena certain
specific restrictions of the geometry and velocities of the system must be
verified. Under these conditions, the role of the Coriolis force can make
the potential asymmetric and favors the orientation of the particles and
therefore their velocities will be in harmony with a vorticity. In the last
part, we study the differential system given by Navier-Stokes and
Ginzburg-Landau equations. Where, \ we obtain a theorem for a modified
Navier-Stokes equations.

\end{document}